\documentclass[apjl]{emulateapj}
\usepackage{rotating}
\usepackage{graphicx}
\usepackage{grffile}
\usepackage{amsmath}
\usepackage{multirow}
\usepackage{longtable}
\usepackage{natbib}
\usepackage{booktabs}
\usepackage{hyperref}
\usepackage{color}

\begin{document}
\setcounter{secnumdepth}{2}
\title{WASP-12\lowercase{b} and HAT-P-8\lowercase{b} are Members of Triple Star Systems}
\author{Eric B. Bechter\altaffilmark{1}, Justin R. Crepp\altaffilmark{1}, Henry Ngo\altaffilmark{2}, Heather A. Knutson\altaffilmark{2}, Konstantin Batygin\altaffilmark{2}, Sasha Hinkley\altaffilmark{3,7}, Philip S. Muirhead\altaffilmark{3,4,8}, John Asher Johnson\altaffilmark{2,5}, Andrew W. Howard\altaffilmark{6}, Benjamin T. Montet\altaffilmark{3,9}, Christopher T. Matthews\altaffilmark{1}, Timothy D. Morton\altaffilmark{3}}
\email{ebechter@nd.edu} 
\altaffiltext{1}{Department of Physics, University of Notre Dame, 225 Nieuwland Science Hall, Notre Dame, IN, 46556, USA}
\altaffiltext{2}{Department of Planetary Science, California Institute of Technology, 1200 E. California Blvd., Pasadena, CA 91125, USA}
\altaffiltext{3}{Department of Astronomy, California Institute of Technology, 1200 E. California Blvd., Pasadena, CA 91125, USA }
\altaffiltext{4}{Institute for Astrophysical Research, Boston University, 725 Commonwealth Ave., Boston, MA  02215 USA}
\altaffiltext{5}{Harvard-Smithsonian Center for Astrophysics, Cambridge, MA 02138, USA}
\altaffiltext{6}{Institute for Astronomy, University of Hawaii, 2680 Woodlawn Drive, Honolulu, HI 96822}
\altaffiltext{7}{NSF Postdoctoral Fellow}
\altaffiltext{8}{Hubble Fellow}
\altaffiltext{9}{NSF Graduate Research Fellow }

\begin{abstract} 

We present high spatial resolution images that demonstrate WASP-12b and HAT-P-8b orbit the primary star of hierarchical triple star systems. In each case, two distant companions with colors and brightness consistent with M dwarfs co-orbit the hot Jupiter planet host as well as one another. Our adaptive optics images spatially resolve the secondary around WASP-12, previously identified by \citet{bergfors_11} and \citet{crossfield_12}, into two distinct sources separated by $84.3\pm0.6$ mas ($21 \pm 3$ AU). We find that the secondary to HAT-P-8, also identified by \citet{bergfors_11}, is in fact composed of two stars separated by $65.3 \pm 0.5$ mas ($15 \pm 1$ AU). Our follow-up observations demonstrate physical association through common proper-motion. HAT-P-8~C has a particularly low mass, which we estimate to be $0.18 \pm 0.02~\mbox{M}_{\odot}$ using photometry. Due to their hierarchy, WASP-12~BC and HAT-P-8~BC will enable the first dynamical mass determination for hot Jupiter stellar companions. These previously well-studied planet hosts now represent higher-order multi-star systems with potentially complex dynamics, underscoring the importance of diffraction-limited imaging and providing additional context for understanding the migrant population of transiting hot Jupiters. 

%Forthcoming observations like those presented here will identify and facilitate the study of systems having experienced Kozai-Lidov oscillations, disk torque interactions or other phenomena.
\end{abstract}

\section{INTRODUCTION}\label{sec:intro} 
There is much debate regarding the origin and evolutionary history of hot Jupiters. Traditional core accretion theory suggests that such planets form beyond the ice-line (the boundary outside which water exists in a frozen state) prior to moving inwards \citep{pollack_96}. The earliest proposed planet migration mechanisms involve a gradual inward-spiral facilitated by planet-disk interactions \citep{goldreich_80, lin_96, murray_98}. Naive interpretation of these migration models presumes that planetary orbits should be well aligned with the spin-axis of their host star. However, precision radial velocity (RV) measurements exploiting the Rossiter-Mclaughlin (RM) effect show that many transiting hot Jupiter orbits are significantly misaligned \citep{winn_09, winn_10b, triaud_10, hebrard_2011, albrecht_12}. 

\begin{figure*}
\begin{center}
\includegraphics[height=2.9in]{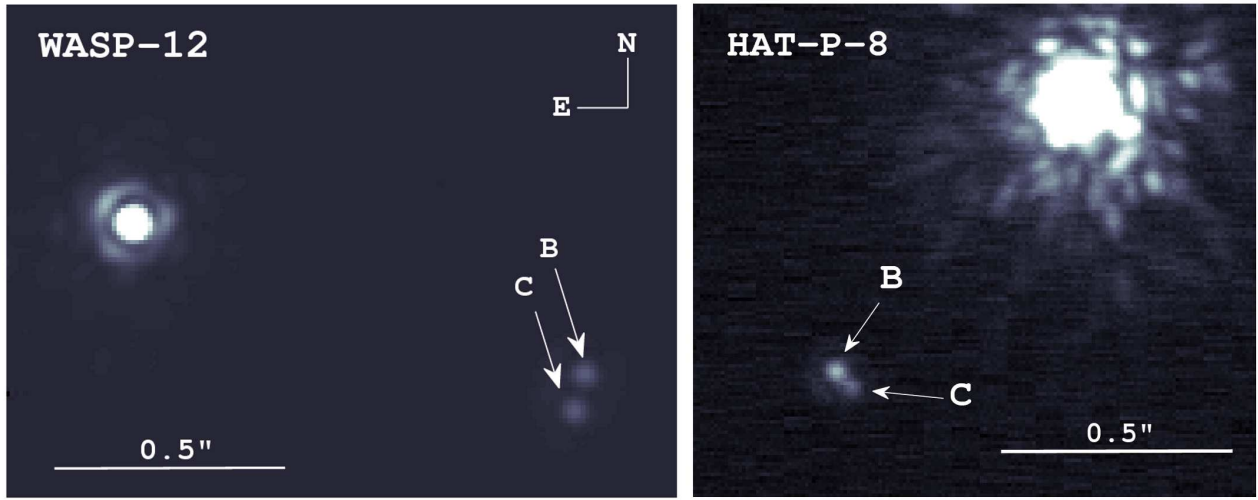} 
\caption{Keck AO discovery images of WASP-12 B,C taken on UT 2012-Feb-02 (left) and HAT-P-8 B,C taken on UT 2012-June-24 (right). North is up and east is left in both images. Follow up observations separated by more than one year recover each companion. } 
\end{center}\label{fig1}
\end{figure*} 

Numerous dynamical models have been proposed to explain the wide range of observed spin-orbit angles, including planet-planet scattering \citep{ford_rasio_08, chatterjee_08} and Kozai-Lidov perturbations with subsequent tidal friction \citep{wu_murray_03, naoz_11}. Several teams have performed comparative analyses suggesting that these two modes could be responsible for placing Jupiters into very short (several day) orbital periods, either individually or in combination \citep{fabrycky_09, morton_johnson_10, nagasawa_11, beauge_12, dawson_13}.  

The underlying assumption motivating these more complex dynamical models is that protoplanetary disks maintain alignment with their host star throughout the planet formation process. However, this assertion need not apply to all stars. Recent theoretical work indicates that the cause of misalignment may instead be induced by forces acting on the disk itself. For example, \citet{lai_11} have proposed that  young protostars with strong magnetic fields ($>$$10^3$ Gauss) can act to warp and misalign the circumstellar disk. Alternatively, gravitational torques from a companion star can change the inclination of the disk relative to the spin-axis of the star prior to the formation of planets \citep{batygin_12}. In any case, such mechanisms must be able to account for the observed abrupt change in the distribution of spin-axis angles as a function of stellar effective temperature \citep{winn_10, albrecht_12}.

A number of plausible hot Jupiter migration mechanisms involve the presence of a massive third body. High spatial resolution imaging can detect such companions at physical scales corresponding to the expected location of their orbits \citep{eggenberger_07, daemgen_09, mugrauer_09, mason_11, roberts_11, ginski_12, faedi_13,  narita_12}.\footnote{Stellar companions at short orbital periods can be constrained and sometimes ruled out by existing RV measurements.} 
%For example, \citet{bergfors_13} have observed a sample of hot Jupiter planet hosts using ``lucky'' imaging and discovered that a large 
These studies consistently find that a significant fraction (tens of percent) host a distant stellar candidate companion that could potentially affect the dynamical histories of the observed hot Jupiters. Several of the most comprehensive and recent programs have used ``lucky'' imaging to efficiently explore a large number of targets. However, near-infrared observations combined with adaptive optics (AO) provides comparatively deeper effective contrast levels especially for objects with red colors such as M dwarfs and brown dwarfs \citep{fleming_12}. 
%Although lucky imaging can efficiently explore
We have recently commenced a multi-faceted observing program, named ``Friends of Hot Jupiters''(hereafter FHJ), that systematically searches for additional companions around a large sample of transiting planet systems \citep{knutson_2013}. The primary objective of the FHJ survey is to quantify the relative fraction of systems, including both well-aligned and misaligned hot Jupiters, that contain distant tertiary bodies, and to study any candidate perturbers in detail using imaging and spectroscopy. 

%Friends began observations at Keck in February 2012.

In this paper, we present initial results from the FHJ survey demonstrating that WASP-12 and HAT-P-8 are actually triple star systems. The candidate companion pairs found orbiting these two planet hosts were identified previously by \citet{bergfors_11} as single objects. Our diffraction limited observations, using Keck, spatially resolve each secondary source into two distinct components. Combining our measurements with previous observations increases the astrometric time baseline by a factor of 2-3 and allows us to confirm the physical association of these objects with their parent star.

\section{SUMMARY OF PREVIOUS OBSERVATIONS}
\subsection{WASP-12}
WASP-12b is a highly irradiated transiting hot Jupiter that orbits a G0V star with a 1.09 day period \citep{hebb_09}. RM measurements yield a sky-projected spin orbit angle of $\lambda$ = 59$^{+15}_{-20}$ deg \citep{albrecht_12}. WASP-12b may have a prolate shape and be undergoing Roche-Lobe overflow that results in substantive mass loss \citep{li_10,fossati_10, fossati_13}. 
%Near-UV observational data of WASP-12 indicate that WASP-12b is indeed experiencing Roche-Lobe overflow \citep{fossati_10, fossati_13}.  
It has been suggested that this planet's dayside emission spectrum is consistent with a super-solar carbon-to-oxygen ratio (\citealt{madhusudhan_11, moses_13}; see however \citealt{crossfield_12}).
%although this interpretation has been debated in the literature \citep{crossfield_12}.
Recent observations of WASP-12b's transmission spectrum indicate that it may also have a high-altitude haze or cloud layer \citep{swain_13, stevenson_13}.

\citet{bergfors_11} detected a faint source separated by $1.047\pm0.021$'' from the WASP-12 primary. Using Keck/NIRSPEC  archival data, \citet{crossfield_12} analyzed the near-infrared spectrum of the candidate companion and found that that it is consistent with an M dwarf. \citet{crossfield_12} also note that the candidate is abnormally bright for an M dwarf if situated at the same distance as the primary. \citet{bergfors_13} find that the companion's point-spread function (PSF) appears to be elongated in two separate epochs, possibly indicating that it is a marginally resolved triple system. 

%\vspace{20mm}
\subsection{HAT-P-8}
HAT-P-8b is a transiting hot Jupiter that orbits an F5V star with a period of 3.07 days  \citep{latham_09}. Initially suspected to have an inflated radius, recent observations by \citet{mancini_13} indicate a higher density than previously reported. \citet{simpson_13} measure a sky-projected spin-orbit angle of $\lambda$ = 15$^{+33}_{-43}$ deg and \citet{moutou_11} find $\lambda$ = -17$^{+9.2}_{-11.5}$ deg, both consistent with a reasonably well-aligned prograde orbit.  High spatial resolution imaging by \citet{bergfors_11,bergfors_13} indicates that HAT-P-8 may be part of a binary star system, although \citet{faedi_13} were unable to confirm the candidate companion, which had a purported angular separation of $1.027 \pm 0.011$". 

%Both groups find that HAT-P-8b moves in a prograde orbit.
\begin{table*}[!ht]
\caption{} Summary of astrometric measurements listing integration time ($\Delta t_{int}$), angular separation ($\rho$), and position angle (PA). Observations are separated by more than one year for each stellar system.
\begin{center}
    \begin{tabular}{cccccccc}
      \hline\hline
      \multirow{2}{*}{Companion}     &   \multirow{2}{*}{JD-2,450,000}    &     \multirow{2}{*}{Date (UT)}& \multicolumn{3}{c}{$\Delta t_{int}$ (s)} &    \multirow{2}{*}{$\rho$ (mas)} &    \multirow{2}{*}{PA ($^{\circ}$)}  \\
 &&&J&K${'}$&K$_{s}$&&\\
\hline\hline
WASP-12~B   &      5,959.9     &      2012-Feb-02      & 135  	&  135&&$1064 \pm 19$      	 & $251.3\pm1.0$ \\
WASP-12~C   &       ---             &            ---                &  ---       &  ---&	&$1073 \pm19$        	& $246.8\pm1.0$ \\
WASP-12~B   &      6,353.8     &      2013-Mar-02      &  	       &    &150	&$1062 \pm18$        	& $251.4\pm1.0$ \\
WASP-12~C   &       ---             &             ---               &   		&    &---	&$1072 \pm18$        	& $247.1\pm1.0$ \\
\hline
HAT-P-8~B     &      6,103.0    &       2012-June-24   &     162	&	95&    	&$1040\pm14$          	&  $137.9\pm0.8$    \\
HAT-P-8~C    &             ---        &              ---              &    ---	&	---&	        &$1049\pm14$          	&  $141.4\pm0.8$    \\
HAT-P-8~B    &       6,476.9    &         2013-July-03    &    180	&	    &180	&$ 1053\pm14$         	&  $137.6\pm0.8$     \\
HAT-P-8~C    &          ---          &               ---               &    ---    &	   &---	&$ 1041\pm14$         	&  $140.7\pm0.8$    \\
\hline
\end{tabular}

  \end{center}
  \end{table*}

\begin{table*}[!ht]
\caption{} 

Secondary and tertiary companion photometric properties. We estimate spectral types using near-infrared color information (when available) and absolute magnitudes by comparing to \citealt{kraus_07}. Absolute magnitudes are found using (photometric or spectroscopic) distance modulus estimates: $d=250\pm30$ pc for WASP-12 \citep{bergfors_13} and $d=230\pm15$ pc for HAT-P-8 \citep{latham_09}.\\[0.1in]
\centerline{
\begin{tabular}{ccccccc}
\hline
\hline
Companion       &    $\Delta J$           &    $\Delta K_{s}$                &    $J-K_s$          &     $M_{K_s}$              & Mass ($M_{\odot}$)   &       Spec. Type  \\
\hline
\hline
WASP-12~B   &  $3.81\pm0.05$     &     $3.25\pm0.04$          &     $0.85\pm0.08$         &     $6.47\pm0.27$      &           $0.38\pm0.05$              &   M3V    \\
WASP-12~C   &  $3.92\pm0.05$    &      $3.28\pm0.04$          &     $0.93\pm0.08$         &     $6.50\pm0.27$      &           $0.37\pm0.05$            &   M3V    \\
\hline
HAT-P-8~B     &     ---       &       $5.58\pm0.07$            &       ---                    &        $7.73\pm0.16$       &          $0.22\pm0.03$         &    $\approx$M5V     \\
HAT-P-8~C     &     ---                        &        $6.08\pm0.10$           &      ---                     &        $8.32\pm0.17$        &         $0.18\pm0.02$          &    $\approx$M6V      \\
\hline
\end{tabular}}
\label{tab2}
\end{table*}
\section{ADAPTIVE OPTICS IMAGING}
We initially observed WASP-12 (V$=11.6$) and HAT-P-8 (V$=10.4$) as part of the FHJ program in Spring 2012 using NIRC2 (instrument PI: Keith Matthews) with the Keck II AO system (Wizinowich 2000). Our standard procedure for searching the immediate vicinity of transiting planet hosts involves executing a three-point dither pattern that facilitates removal of instrument and sky background radiation while avoiding the (noisy) bottom-left quadrant of the NIRC2 array. Observations are nominally obtained in position angle mode without allowing for field rotation since we do not perform PSF subtraction. We used the NIRC2 narrow camera setting to provide fine (10 mas) spatial sampling of the instrument PSF. Integration times for all observations are listed in Table 1.

The data were processed using standard techniques to flat-field the array, replace hot pixels, subtract background noise, and align and co-add the frames (e.g., \citealt{crepp_12b}). Figure \ref{fig1} shows the final reduced K-band images for WASP-12 and HAT-P-8. Our observations provide a spatial resolution comparable to the diffraction limit (approximately 45 mas). In each case, two candidate companions (B, C) are detected. 

We obtained complementary photometry in the J-band to determine the companion colors and help constrain their physical properties. WASP-12~BC are spatially resolved in the J-band; however HAT-P-8~BC are not seen in the UT 2012 June-24 J-band images due to high airmass (2.19) indicating that the HAT-P-8 companions have red colors. Deeper follow-up J-band observations taken UT 2012 July-03 (see Section 4.2) detect the combined light of HAT-P-8~BC but do not spatially separate the objects as is seen at longer wavelengths.

\begin{figure*}
\begin{center}
\includegraphics[height=2.9in]{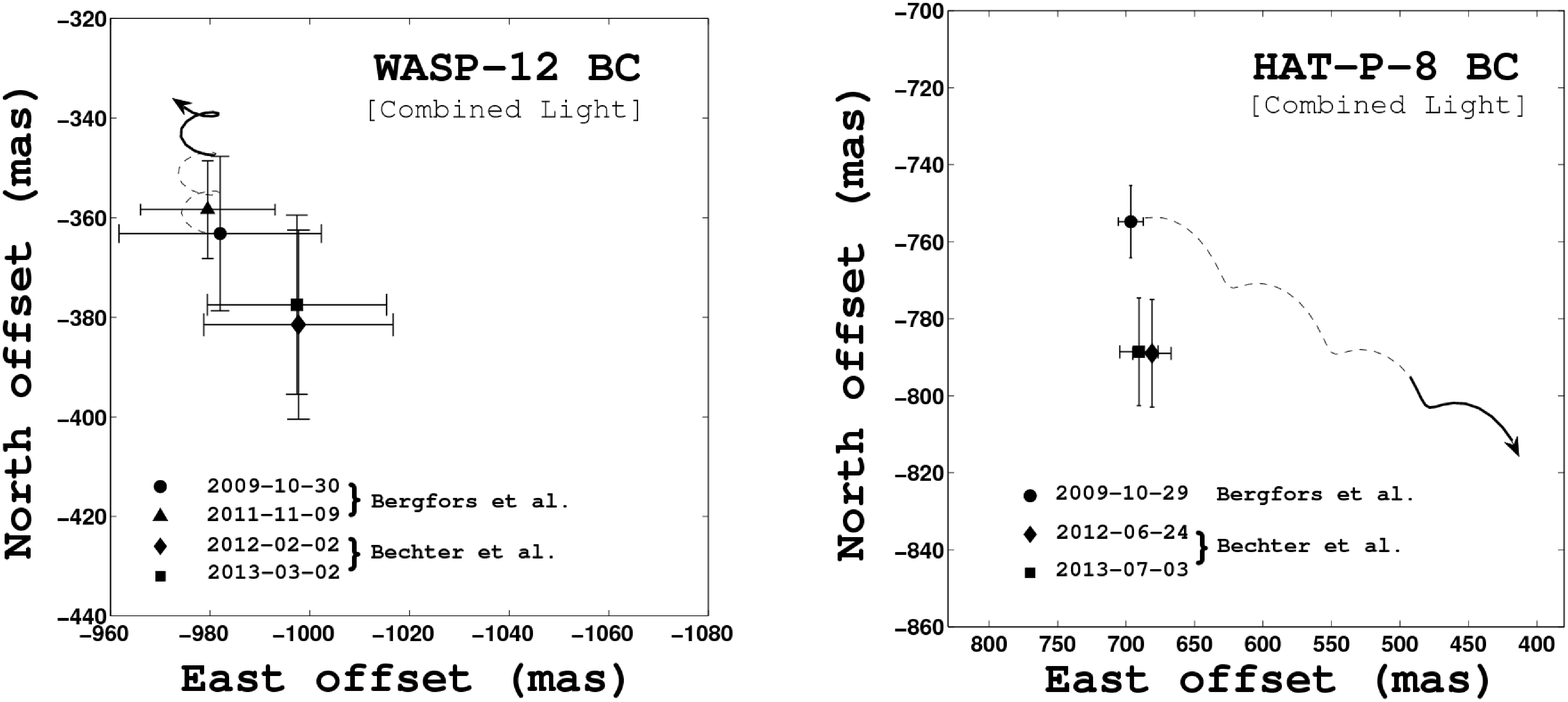} 
\caption{Astrometric measurements for WASP-12 and HAT-P-8. Axes correspond to the angular separation (offset) in the north and east cardinal directions as measured relative to the primary star. The combined proper motion plus parallactic motion of an infinitely distant (unassociated) object is given by the dashed and solid curves. Dashed curves correspond to the astrometric time baseline of \citet{bergfors_13}. The solid curves correspond to the astrometric time baseline of this study. \citet{bergfors_13} did not spatially resolve the WASP-12~BC and HAT-P-8~BC components, but did provide the initial detection of their combined light signal (in 2009 October). We plot  the photo-center of our resolved BC companions to compare to Bergfors' 2009 and 2011 data. Our Keck AO epochs are separated by more than one year and demonstrate physical association (by themselves) for HAT-P-8; association of WASP-12 BC is established by combining our results with those of \citet{bergfors_13}. Our astrometric uncertainties are over-plotted and comparable to \citet{bergfors_13}. Our measurement precision is dominated by systematics from distortion in the individual frames. Orbital motion of these two bodies may be detectable with additional observations.}  
\end{center}\label{fig2}
\end{figure*} 

\section{PHOTOMETRY AND ASTROMETRY} 
\subsection{PSF Model Fits}
We perform a Bayesian analysis to model the AO observations of WASP-12 and HAT-P-8 at each epoch. Specifically, Markov-Chain Monte Carlo (MCMC) numerical methods are employed to compute companion relative brightnesses, astrometric positions, and determine uncertainties. The Metropolis-Hastings algorithm efficiently explores regions of parameter space to find the best-fitting global minimum and calculate posterior distributions for each fit parameter. 

We simultaneously model three point-spread functions to self-consistently account for contamination from nearby companion star. Free parameters include: rectilinear coordinates for each source; peak brightness of each source; sky background levels (which we model as spatially uniform); and PSF fitting parameters, $\alpha$, $\beta$, $\gamma$, $r_s$, and $w$. The observations are well-modeled using a modified Moffat function, given by: 
\begin{equation}
I(x,y)= \sum_{i=1}^{i=3}\: \left\{ \alpha_i \left[1+\left(\frac{r_i}{r_s}\right)^{2}\right]^{-\beta}+\gamma_i\:e^{-r_i^2/w^2} \right\}, 
\end{equation}
where $r_i= \sqrt{(x_i-{x_0}_i)^2+(y_i-{y_0}_i)^2}$ is a polar coordinate corresponding to the angular separation from each source, $i$, in the image. The term on the left describes the AO halo and the term on the right characterizes the PSF core. By separating the terms, we effectively account for tip/tilt and focal anisoplanatism in the images, although we do not allow $r_s$ and $w$ to vary individually ($w_i = w$, $r_{s{_i} }= r $) due to the already large number of degrees of freedom (twelve when including the sky background). The posterior distributions found by our MCMC algorithm marginalize over all fitting parameters. 

Equation 1 captures on-axis AO features but does not replicate low order aberrations or diffraction from the first Airy ring. We have experimented with other PSF forms such as sinc(...) and sinc$^2$(...) functions. Assuming that uncertainties in each reduced image are described by Poisson statistics at the pixel level, resulting from sky-background subtraction shot-noise, we find the results from each AO model are consistent with one another but uncertainties are unrealistically small. For example, angular separation measurement uncertainties are less than 1 mas (1$\sigma$). The images used for our analysis have been fully processed prior to MCMC calculations. As such, we have stacked frames acquired from different dither positions. This step is required because the companions are so much fainter than their primary star particularly in the J-band. However, by combining images obtained from different locations on the array, we have introduced PSF spatial smearing from uncorrected optical distortions. We estimate the size of this effect using polynomial fits available for the NIRC2 array provided by Keck Observatory\footnote{Distortion correction polynomials found \href{http://www2.keck.hawaii.edu/inst/nirc2/dewarp.html}{\textcolor{blue}{here}.}}. Systematic errors are of order 1-2 pixels and change depending on the size of the dither pattern. Distortion corrections may be applied before image stacking but this introduces significant numerical noise. Furthermore, the correction coefficients also change slowly with time \citep{yelda_2010}.

Our final adopted astrometric uncertainties were found by adding the effects of optical distortion in quadrature with that from photon noise and pixel crosstalk resulting from PSF fitting errors. We self-consistently account for uncertainty in the plate scale and orientation of the NIRC2 array \citep{ghez_08} by randomly drawing values for the plate-scale and orientation from a normal distribution and folding the results into calculations of the angular separation and position angle when converting from pixel separations to arcseconds. Nevertheless, the effect from optical distortion dominates the uncertainty for each astrometric epoch as it is much greater than both pixel cross-talk and photon noise. Results for relative astrometry measurements are shown in Table 1. Although our observations from 2012 and 2013 were acquired in different filters ($K'$ and $K_s$), due to a change in the FHJ default observing strategy, this effect appears to be small since the results are nearly identical. 

%The starlight takes a different path through the optics creating systematic errors that are larger than the apparent astrometric jitter caused by photon noise. 

%This effect is however degenerate with any possible orbital motion. 

%In order to compare our results to \citet{bergfors_13} we calculate the combined light centroid using a weighted average of WASP-12 BC and HAT-P-8 BC.
%Our results are compared to Bergfors in Figure 2.
%To calculate our final astrometric uncertainty shown in Figure 2, we estimate the errors from a number of sources including: distortion, empirical fit uncertainty, and sky-background noise. When reducing our data, we cannot apply a distortion correction for NIRC2 as our B and C companions are not visible in several frames prior to stacking. To estimate this distortion, we use the correction polynomial available for the NIRC2 array to obtain the correction factor for each dither point. From this we find the distortion error to be $~$2 pixels. Using an empirical model (Moffat function), instead of the primary star PSF, to fit the secondary and tertiary introduces a small astrometric uncertainty ($<0.5$ pixels). Additionally, the systematic error introduced by subtracting sky-background is set by Poisson noise ($0.1$ pixels). Thus, we find our astrometric uncertainty to be dominated by distortion (2 pixels).

\begin{figure*}
\begin{center}
\includegraphics[height=2.8in]{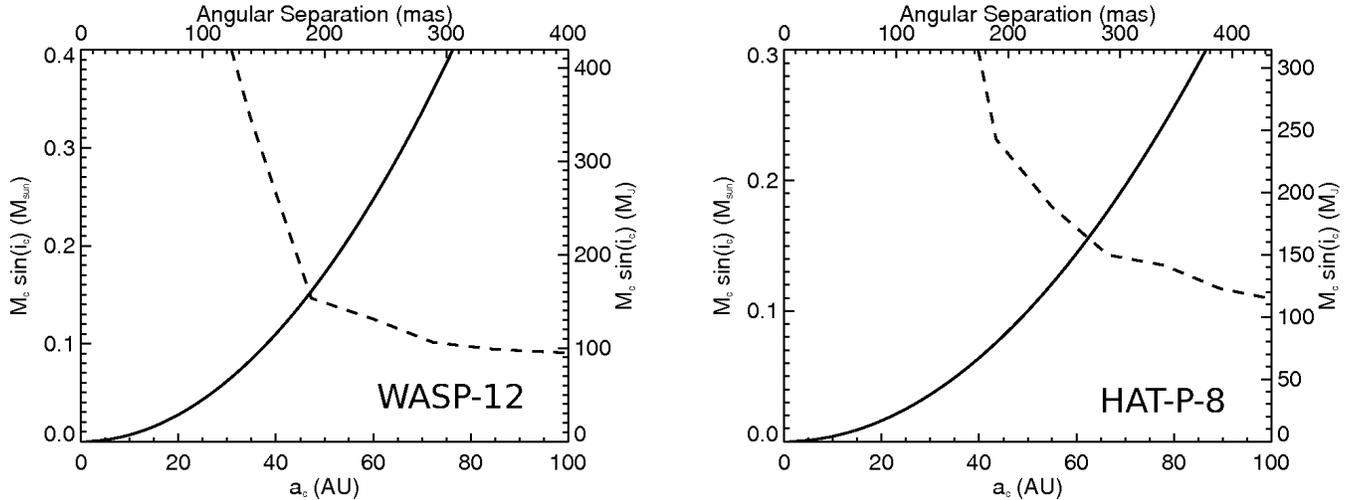} 
\caption{Joint RV and imaging constraints on the presence of additional companions orbiting WASP-12 (left) and HAT-P-8 (right) using the accelerations listed in Equation 5. Any unseen stars, brown-dwarfs, or gas giant planets must lie below both the limits set by Doppler RV measurements (solid line) and those set by AO imaging (dashed line). } 
%Uncertainty in the location of additional putative companions is limited by uncertainty in the distance to each star ( $2 \sigma$ limits are given by dashed lines). On-sky contrast curves are over-plotted for reference as well as the location of WASP-12~BC and HAT-P-8~BC (note: vertical axis is a minimum mass). }
\end{center}\label{fig3}
\end{figure*} 
\subsection{Physical Association}
We perform an astrometric analysis to assess the physical association of the WASP-12~BC and HAT-P-8~BC candidates with their primary star. To do so, we compare our astrometric measurements against the null hypothesis that the off axis sources are infinitely distant unrelated background objects with zero parallax. WASP-12 has a small proper motion [-0.7, -7.8 mas/yr] comparable to the size (9.963 $\pm$ 0.006 mas) of a NIRC2 pixel \citep{hog_00, ghez_08}. HAT-P-8 has a proper motion that is an order of magnitude larger [75.5, 17.5 mas/yr] \citep{hog_00}. Neither star has a \textit{Hipparcos} parallax measurement which complicates the analysis. Instead, the distance to WASP-12 is estimated using a photometric distance modulus \citep{bergfors_13} and the distance to HAT-P-8 is determined using a spectroscopic distance modulus \citep{latham_09}. We incorporate parallactic motion by converting estimated distances to a trigonometric parallax (ellipse). The resulting differential motion across the sky between the primary star and candidate secondary/tertiary is given by the vector sum of the proper and parallactic motion \citep{zimmerman_10}. 

Our astrometric measurements are shown in Figure 2. Over-plotted are previous measurements taken by \citet{bergfors_13} in October 2009 that identify the combined light signal of WASP-12~BC and HAT-P-8~BC but do not spatially resolve the sources into individual components. Our Keck AO observations from 2012 and 2013 clearly separate the light from each companion star. The angular separation of WASP-12~BC is $84.3\pm0.6$ mas ($21\pm3$ AU) and the angular separation of HAT-P-8~BC is only $65.3\pm0.5$ mas ($15\pm1$AU), comparable to the diffraction limit of a 10 meter telescope at near-infrared wavelengths. Optical distortion for such small separations is negligible. To compare data on an equal footing with \citet{bergfors_13}, we plot combined light photo-centers for WASP-12 BC and HAT-P-8 BC in Figure 2. 

The \textit{a priori} probability of finding three point sources in a hierarchical configuration separated by only 1'' on the sky is very low. Our two astrometric epochs for WASP-12 and HAT-P-8 are separated by 393.9 days and 373.9 days, respectively. The expected motion of a background source relative to the primary star is $8.5 \pm 1.0$ mas ($0.9 \pm0.1$ pixels) for WASP-12 and $79.3\pm2.9$ mas ($8.0\pm0.3$ pixels) for HAT-P-8 over the same time-frame. 

%Given the brightness of each source and our integration time, the astrometric precision set by photon noise is approximately $0.1$ pixels for each epoch.

With only two observations, the confirmation that WASP-12~BC are bona-fide companions is marginal. However, combining our measurements with the 2009 October initial detection from \citet{bergfors_11} we can demonstrate that the three point sources are physically associated (Figure 2). To further reinforce our results, we have determined the photometric distance modulus for WASP-12 BC. The combined light apparent magnitude of the WASP-12 System is 10.19$\pm0.02$ \citep{skrutskie_06}. Backing out the individual apparent magnitudes of WASP-12 BC from our relative photometry measurements, we find the distance to WASP-12 B is 263$\pm13$ pc and the distance to WASP-12 C is 267$\pm13$ pc. These values overlap with the photometric distance estimated by \citet{bergfors_13} of 250$\pm30$ pc ruling out the possibility that they are foreground or background objects. 

%Additionally, we calculate photometric distances to WASP-12 B and C to rule out the possibility of distant giant stars and near-field dwarf stars. Using an estimated spectral type of M3 for WASP-12 B \& C (Table 2), we find them to be at  and ..., while the primary star is at a distance of 250 $\pm30$ pc, further suggesting they are physically associated with WASP-12 A. 

HAT-P-8~BC are confirmed using our observations alone due to the large space motion of the host star. We cannot claim detection of orbital motion for either system because of the aforementioned systematic errors and fact that the stars were observed with different instruments and filters. Dedicated astrometric measurements are required to determine the total dynamical mass of the secondary and tertiary in each case \citep{dupuy_10}. We note that in both cases, WASP-12 and HAT-P-8, our measurements are $\approx$ 20 mas south and $\approx$ 20 mas east of Bergfors suggesting possible systematics between the AstraLux and Keck data sets.

\subsection{Companion Characterization} 

\citet{bergfors_13} assign a preliminary spectral type of M0V for WASP-12~``B'' (combined light), assuming the identified off axis source is associated. \citet{crossfield_12} find that WASP-12~``B'' is a hot M dwarf with $\Delta K=2.45\pm0.06$ mag. We find that WASP-12~B and WASP-12~C are $\Delta K^{A,B}_s = 3.25\pm0.04$ and $\Delta K^{A,C}_s = 3.28 \pm 0.04$ mags fainter than the primary respectively (Table 2). Combining the signal from both components, our measurements show that the expected unresolved brightness difference between the secondary/tertiary and primary star should be $\Delta K^{A,BC}_s = 2.51\pm0.03$ mag, consistent with the interpretation of \citet{crossfield_12}.

To further characterize the companions around each star, we calculate absolute magnitudes based on previous distance estimates from \citet{bergfors_13} and \citet{latham_09}. Our uncertainty in absolute magnitude is dominated by the lack of a trigonometric parallax measurement. We estimate the mass of each companion using \cite{girardi_02} evolutionary models assuming a system age of 5 Gyr. Comparing our absolute magnitudes to those of \citet{kraus_07}, we find that WASP-12~BC are consistent with M3V (Table 2). Additionally, the J - K colors of WASP-12~BC (see Table 2) are also consistent with M stars \citep{kraus_07}. Although HAT-P-8~BC are detected during second epoch (UT 2013 July-03) observations, they are spatially unresolved in the J-band because the images were obtained at an airmass of 2.19. Performing aperture photometry for the pair, we find a combined difference in magnitude of $\Delta J^{A,BC} J=5.9\pm0.2$. We estimate the spectral types of HAT-P-8 B and C to be $\approx$ M5V and M6V respectively using K-band photometry alone. 
\subsection{Companion Constraints}
As part of the FHJ program, we obtained additional RV measurements for both systems, which we use to constrain the presence of additional companions at shorter orbital periods. Our best-fit RV slopes are: 

\begin{eqnarray}
{dv/dt}_{WASP-12}= -4.12 \pm 4.37~\mbox{m/s/year} \nonumber \\
{dv/dt}_{HAT-P-8}= -2.72 \pm 2.39~\mbox{m/s/year} ,
\end{eqnarray} 

consistent with the absence of massive, $m\geq5~M_J$, objects out to $a \leq 8.3$ AU for WASP-12 and $a \leq10.9$ AU for HAT-P-8. Figure 3 displays joint constraints imposed by the combination of Doppler RV measurements (solid lines) and direct imaging observations (dashed lines). Should any additional companions be present in these systems, their masses must reside below both curves. Continued RV monitoring of the host stars will further eliminate regions of mass-semi-major axis parameter space.

\section{SUMMARY \& DISCUSSION}

We have commenced a multi-disciplinary follow-up observing program, named ``Friends of Hot Jupiters'' (FHJ), that targets a large sample of short-period gas giant transiting planet systems. In this paper, we present AO images from Keck that spatially resolve previously identified candidate companions around WASP-12 and HAT-P-8 into two distinct sources. When combined with previous observations from \citealt{bergfors_13}, our astrometric measurements show that WASP-12~BC and HAT-P-8~BC are gravitationally bound to one another as well as the primary, making WASP-12b and HAT-P-8b members of hierarchical triple star systems. 

%We used a Bayesian analysis method to model the AO observations at each epoch along with MCMC simulations to determine relative brightnesses and astrometric positions. 

Our diffraction-limited measurements show that the two companions around WASP-12 are separated by $84.3\pm0.6$ mas ($21\pm3$ AU) and have roughly equal brightness. We estimate spectral types of M3V, consistent with the \citealt{crossfield_12} (spatially unresolved) combined-light spectroscopic analysis. Our photometric measurements combined with evolutionary models indicate masses of $0.38\pm0.05~\mbox{M}_{\odot}$ and $0.37\pm0.05~\mbox{M}_{\odot}$ for WASP-12 B and C, respectively. The companions orbiting HAT-P-8 are separated by only $65.3 \pm0.5$ mas ($15\pm1$ AU) and have somewhat more disparate properties. We estimate that HAT-P-8~B has a mass of $0.22\pm0.03~\mbox{M}_{\odot}$ and HAT-P-8~C has a mass of $0.18\pm0.02~\mbox{M}_{\odot}$. In each case our ability to characterize each system is limited by the lack of an accurate trigonometric parallax. 

% The distance to HAT-P-8 is estimated to be $230\pm15$pc by \citet{latham_09}, however \citet{bergfors_13} calculate a distance to HAT-P-8 of only $150\pm20$ pc or $190\pm20$ pc  as they found equally good fits for spectral types F8V and F5V. This discrepancy in distance limits our our ability to characterize each system, underlining the importance for accurate trigonometric parallax measurements. 

%(?)The companion around WASP-12 was originally characterized as an unresolved $\approx$M0V star \citep{crossfield_12, bergfors_13}. 

%Observations with the Hubble Space Telescope  spectral types of $\approx$M

%The ongoing debate concerning the origin of misaligned hot Jupiters has brought about several potential orbital evolutionary theories including: planet-planet scattering \citep{ford_rasio_08, chatterjee_08}, Kozai-Lidov perturbations with tidal friction \citep{wu_murray_03, naoz_11}, and disk torque mechanisms \citep{batygin_12}. 

The ongoing debate concerning the origin of misaligned hot Jupiters has brought about several potential orbital evolutionary theories. AO imaging shows significant promise to improve our understanding of the dynamical history of these systems. Although numerous candidate companions around hot Jupiter hosts have been identified (e.g., \citealt{bergfors_13}), multi-epoch astrometry that assesses the physical association of these objects requires dedicated follow-up measurements from comprehensive programs that study close-seperations stellar companions in detail. WASP-12 and HAT-P-8 may offer unique insights into the dynamics of hot Jupiter systems because their hierarchy will ultimately enable companion mass estimates using dynamics.

%\citet{winn_10} finds a dichotomy (in sky-projected spin orbit angles) $\lambda$ angels that indicates hot stars (>6250 K) may be responsible for the misalignment of hot Jupiters. Interestingly, WASP-12 fits the stellar temperature criteria (6250K) that \citet{winn_10} finds associated with highly oblique hot Jupiters.

%(must exercise caution when attempting a dynamical mass calculation as observations used two different instruments.)

\section{ACKNOWLEDGEMENTS}
This research has made use of the SIMBAD database, operated at CDS, Strasbourg, France. Data presented herein were obtained at the W.M. Keck Observatory, which is operated as a scientific partnership among the California Institute of Technology, the University of California and the National Aeronautics and Space Administration. The Observatory was made possible by the generous financial support of the W.M. Keck Foundation. J.A.J. is supported by generous grants from the David and Lucile Packard Foundation and the Alfred P. Sloan Foundation. B. T. M. is supported by the National Science Foundation Graduate Research Fellowship under Grant No. DGE1144469.

\bibliography{ms}

\end{document}